\documentclass[10pt,a4paper]{article}
\usepackage[a4paper,margin=1in]{geometry}
\usepackage{graphicx, multirow, outlines, ulem}
\usepackage[latin1]{inputenc}
\usepackage{amsmath} 
\DeclareMathOperator{\sech}{sech}
\DeclareMathOperator{\cosech}{cosech}
\usepackage{amsfonts}
\usepackage{amssymb}
\usepackage{caption}
\usepackage{subcaption}
\usepackage{float}
\usepackage{cite}
\usepackage{epstopdf}
\DeclareGraphicsExtensions{.pdf,.png,.jpg}
\usepackage{breqn}
\usepackage{upgreek}
\makeatother
\makeatletter
\newcommand{\mathleft}{\@fleqntrue\@mathmargin0pt}
\newcommand{\mathcenter}{\@fleqnfalse}
\usepackage{xcolor}
\usepackage[colorlinks,citecolor=red,linkcolor=blue]{hyperref}
\def\beq{\begin{equation}}
\def\eeq{\end{equation}}
\def\bea{\begin{eqnarray}}
\def\eea{\end{eqnarray}}

\begin{document}

\begin{center}
  {\Large \bf  Generating QES potentials supporting zero energy normalizable states for an extended class of truncated Calogero Sutherland model}
\vspace{1.3cm}

{\sf Satish Yadav\footnote[1]{e-mail address:\ \ s30mux@gmail.com,\hspace{0.05cm} Satish.yadav17@bhu.ac.in},
Sudhanshu Shekhar \footnote[2]{e-mail address:\ \ sudhanshu251997@gmail.com,\hspace{0.05cm}sudhanshushekhar@bhu.ac.in },
Bijan Bagchi \footnote[3]{e-mail address:\ \ bbagchi123@gmail.com,\hspace{0.05cm} },
Bhabani Prasad Mandal \footnote[4]{e-mail address:\ \ bhabani.mandal@gmail.com,\hspace{0.05cm} bhabani@bhu.ac.in},
}

\bigskip 

{\it $^{1,2,4}$ Department of Physics,
Banaras Hindu University, \\ Varanasi-221005, INDIA. \\ \it $^{3}$ Brainware University, Barasat, \\ Kolkata 700125, INDIA
}

\bigskip
	\noindent 
 
 {\bf\large Abstract}\\

\end{center}

Motivated by recent interest in the search for generating potentials for which the underlying Schr\"{o}dinger equation is solvable, we report in the recent work several situations when a zero-energy state becomes bound depending on certain restrictions on the coupling constants that define the potential. In this regard, we present evidence of the existence of regular zero-energy normalizable solutions for a system of quasi-exactly solvable (QES) potentials that correspond to the rationally extended many-body truncated Calogero-Sutherland (TCS) model. Our procedure is based upon the use of the standard potential group approach with an underlying $so(2, 1)$ structure that utilizes a point canonical transformation with three distinct types of potentials emerging having the same eigenvalues while their common properties are subjected to the evaluation of the relevant wave functions. These cases are treated individually by suitably restricting the coupling parameters.

\medskip
\vspace{1in}

\section{Introduction}  

A major difficulty in pinning down zero-energy quantal states lies in the extraction of corresponding normalizable solutions of the Schr\"{o}dinger equation. This has motivated a lot of research in the look out for the existence of such zero-energy states with valid asymptotic properties \cite{ito1970partantipart, daboul1994quantum,daboul1995exact,daboul1996exact,Alhaidari,Halberg,Makowski,Kaleta}. Interestingly, it was noted \cite{barut1980magnetic} long ago that zero-energy solutions could exist with the properties of confinement (in case of normalizablity)
and leakage (if the wavefunction turns out to be non-normalizable) for certain types of interacting systems. Another example is exemplified in the Coulomb problem, where a zero-energy state is non-normalizable and thus considered a part of the continuum. However, such a situation is not without its exceptions. Indeed, there exist cases pertaining to a QES system where the $E = 0$ state becomes bound should we restrict to certain values of the coupling constant \cite{bagchi1997zero}. 

In a broader context, searching for solvable potentials, that possess an exact spectrum with accompanying closed-form solutions for the wavefunctions, has always been in the spotlight of active research (see, for instance, \cite{shifmanqes, khare1995, fernandez}). Among different techniques that have been adopted in this regard, the ones based on point canonical transformation \cite{levai1989} coupled with employing group theoretical techniques that exploit the structure of the Casimir operator \cite{alhassid1, alhassid2, alhassid3, quesne1991, yadav2015group, ramos2017short}, have evinced most attention. On the other hand, the limited number of exactly solvable (ES) potentials in quantum mechanics has prompted investigation into the question of the existence of quasi-exactly solvable (QES) \cite{turbinerqes1, turbinerqes2, shifmanqes, ushveridzebook, bender1996, tkachuk, brihaye, fring2019, khare1998quasi, khare2009new, basu2017quasi} and conditionally exactly solvable (CES) systems \cite{dutraces1, dutraces2, grosche1, grosche2, duttces, znojil1, znojil2}. Note that while the former types allow for a partial or complete assessment of the energy spectrum under possible constraints among the potential parameters, the latter ones are typical in the sense that one or more
coupling constants present in them need to be tuned to a specific value thus facilitating a valid asymptotic behaviour. Beyond being of academic interest, both QES and CES potentials provide valuable insights in the characterization of physical phenomena. In particular, a connection between QES and  spin-boson and spin-spin interacting systems is well known \cite{zaslavskii1990qes}, and the dual partnership between CES and the accompanying shape invariant ES potential was pointed out some time ago \cite{bagchi2004ces}. 


In this work, we present a novel evidence of zero-energy normalizable solutions for a class of QES rational potentials which correspond to rationally extended many-body truncated Calogero-Sutherland model (TCS) \cite{pittman2017truncated}. Towards such a pursuit, we make use of the algebraic techniques based upon the profitable employment of so$(2,1)$ as a potential algebra for the Schr\"{o}dinger equation. A distinguishing feature of our approach is that for all the three categories of solutions furnished by the latter we are led to regular normalizable wavefunctions provided certain convergence condition is satisfied. The structure of this paper is organized as follows: Section $2$ provides a concise review of the so$(2,1)$ potential algebra discussing its basic layout. In Section $3$, we introduce the rationally extended many-body truncated Calogero-Sutherland model (TCS) while in Section $4$ we utilize the relevance of so$(2,1)$ to determine the zero-energy normalizable solutions for a specific class of QES rational potentials within TCS, subject to certain restrictions on the potential parameters. Finally, in Section $5$, a summary is presented.

\section{so (2, 1) potential algebra}

The underlying commutation relations of so(2,1), namely $[J_+, J_-] = -2J_0$, $[J_0,
J_{\pm}] = \pm J_{\pm}$, are guided by the following differential realizations of the generators
\begin{equation}\label{eq:generators}
J_0 = -i \frac{\partial}{\partial \phi}, \quad
J_{\pm} = e^{\pm i\phi}\left [ \frac{\partial}{\partial x} + F(x) \left ( i \frac{\partial}{\partial \phi} \mp \frac{1}{2} \right ) + G(x) \right ],
\end{equation}
where the two functions~$F$ and~$G$ are controlled by a set of coupled equations
\begin{equation}
  F' = 1 - F^2, \qquad \mbox{and} \qquad G' = - F G,     \label{eq:so(2,1)-cond}
\end{equation}
in which the dashes denoting derivatives with respect to $x$. For bound states, to which we restrict here, one considers unitary irreducible
representations of the $D_k^+$-type that point to basis states which correspond to the eigenfunctions of different Hamiltonians, but conforming to the same energy level. Note that the Casimir operator being $J^2 = J_0^2
\mp J_0 - J_{\pm} J_{\mp}$, it has an explicit form $J^2 = \left(\partial^2/ \partial x^2
\right) - F' \left[\left(\partial^2/ \partial \phi^2\right) + \frac{1}{4}
\right] + 2 i G' \left(\partial/ \partial \phi\right) - G^2 - \frac{1}{4}$.
Hence,  with $J_0 |km\rangle = m |km\rangle$,
and $J^2 |km\rangle = k(k-1) |km\rangle$ ($m=k$, $k+1$, $k+2$,~$\ldots$), where, in the realization of ~(\ref{eq:generators}), the states given by
$|km\rangle = \psi_{km}(x) e^{im\phi}$ are the basis functions, it readily follows that the
coefficient functions $\psi_{km}(x)$ satisfy the Schr\"odinger equation
\begin{equation}
  - \psi_{km}'' + V_m \psi_{km} = - \left(k - \frac{1}{2}\right)^2 \psi_{km}
\end{equation}
in the presence of a one-parameter family of potentials expressed in terms of $F(x)$ and $G(x)$
\begin{equation}
  V_m = \left(\frac{1}{4} - m^2\right) F' + 2m G' + G^2.   \label{eq:V_m}
\end{equation}

%
%
As shown by Wu and Alhassid \cite{alhassid1, alhassid2, alhassid3}, a particular choice of solutions
to ~(\ref{eq:so(2,1)-cond}) correspond to Morse, P\"oschl-Teller, and Rosen-Morse
potentials. Later, an extended formalism by Englefield and
Quesne \cite{quesne1991} identified more general possibilities to include additional solutions according to whether $F^2 < 1$, $F^2 = 1$
or $F^2 > 1$ namely,

\begin{align}\label{eq:3cases}
    \mbox{(I)} \quad  F(x) &= \tanh{(x)}, \qquad  G(x) = b \sech{(x)},\\
    \mbox{(II)} \quad F(x)&= \pm 1, \qquad ~~~~~~G(x) = b e^{\mp x},\label{eq:3cases2}\\
    \mbox{(III)} \quad F(x)&= \coth{(x)}, \qquad G(x) = b \cosech{(x)}\label{eq:3cases3}.
\end{align}

By substituting these solutions into equation (\ref{eq:V_m}) one arrives at  non-singular
Gendenshtein, Morse, and singular Gendenshtein potentials. These solutions
encompass those obtained by Wu and Alhassid in that Gendenshtein
potentials are disguised versions of P\"oschl-Teller potentials; further, for
particular values of the parameters, one gets non-singular Rosen-Morse potentials
from singular Gendenshtein potentials.\par
%
%
\section{Extended truncated Calogero-Sutherland model}
  The extended $N$-body TCS model in one-dimension is concerned with $N$ particles which are harmonically confined and interact
pairwise via two-body potential as well in the presence of a three-body term. The idea of truncation of the interaction refers to the number of neighbours instead of the relative distance of the particles. The relevant  Hamiltonian is characterized by the form 
  \begin{equation}\label{Hamiltonian}
      H_{ext}= \hat{H}+V_{new}
  \end{equation}
  where following the notations of \cite{pittman2017truncated, yadav2019rationally}
 \begin{equation}\label{tcm}
\hat{H}=\sum^{N}_{i=1}\bigg[-\frac{1}{2}\frac{\partial^2}{\partial x^2_i}+\frac{1}{2}\omega^2x^2_i\bigg]+V_{int},
\end{equation}  
\begin{equation}\label{vi}
V_{int}=\sum_{\substack {i<j \\ \mid i-j\mid \le r}}\frac{\lambda (\lambda-1)}{\mid x_i-x_j\mid ^2}+\sum_{\substack {i<j<k \\ \mid i-j\mid \le r\\ \mid j-k\mid \le r}}\frac{\lambda^2 {\bf r}_{ij}.{\bf r_{jk} }}{r^2_{ji}r^2_{jk}};\qquad \lambda \ne 0,
\end{equation}
where ${\bf r}_{ij}=(x_i-x_j)\hat x$ is vector along x-axis, $r$ is the range of interaction  and $V_{new}$ is a new interaction term as given by 
\begin{equation}\label{vnew}
    V_{new}=\frac{(\alpha_1+\alpha_2 \omega^2 \rho^2)}{(\beta_1+\beta_2 \omega^2 \rho^2)^2}
\end{equation}
containing some unknown coefficients, where $\rho^2=\sum^N_{i=1} x^2_i$. The two body interactions in the above scheme are attractive in the range $0<\lambda <1$, and repulsive for $\lambda \geq 1$. It needs to be pointed out that for the specific case of
$r = 1$, the corresponding form of the Hamiltonian goes over to the model of Jain and Khare \cite{Jain}, while for the case $r = N-1$, one recovers what was initially proposed by Calogero\cite{Calogero}, and Sutherland \cite{Sutherland}.  $\rho$ in Eq.  \eqref{vnew} varies from $0 $ to $\infty $  unlike the variable
$x$ in $so(2,1) $ potential algebra, which varies from $-\infty $ to $\infty$.

To find the  solution of the Schr\"{o}dinger equation accompanying Eq. (\ref{Hamiltonian})
\begin{equation}\label{sch}
    \hat{H}_{ext}\Psi_{ext}=E_{ext}\Psi_{ext}
\end{equation}
we consider  $\Psi_{ext}({\bf x})=\phi({\bf x})\xi_{ext}$ with ${\bf x}=(x_1,x_2,....,x_N) \in \mathbb{R}^N$, where 
$ \phi({\bf x}) $ is given by $\Pi_{i<j} (x_i - x_j)^\lambda$, and  $\xi_{ext}$ satisfying 
\begin{equation}
    -\frac{1}{2}\sum^{N}_{i=1}\frac{\partial^2 \xi_{ext}}{\partial x^2_i}-\lambda\sum^{N-1}_{i<j}\frac{1}{x_i-x_j}\bigg(\frac{\partial \xi_{ext}}{\partial x_i}-\frac{\partial \xi_{ext}}{\partial x_j}\bigg)+\bigg(\frac{1}{2}\sum^N_i\omega^2x^2_i+V_{new}-E_{ext}\bigg)\xi_{ext}=0.
\end{equation}
Next, redefining the function $\xi_{ext}=\Phi_{ext}(\rho) P_s({\bf x})$, we find that $\Phi_{ext}(\rho)$ obeys the radial equation \cite{yadav2019rationally} 
\begin{equation}\label{esrho}
    \Phi_{ext}''(\rho)+\big(N+2s-1+ \lambda r(2N-r-1)\big)\frac{1}{\rho}\Phi_{ext}'(\rho)+2\big(E-(\frac{1}{2}\omega^2\rho^2+V_{new})\big )\Phi_{ext}(\rho)=0
\end{equation}
with $P_s({\bf x})$ given by the generalized Laplace equation. Note that in (\ref{esrho}) a prime on $\Phi_{ext}(\rho)$ indicates 
derivative with respect to $\rho$.

To solve it we follow the standard procedure of implementing a point canonical
transformation by substituting\cite{Sudarshan} of substituting 
\begin{equation}\label{extsol}
\Phi_{ext}(\rho)=f(\rho)\zeta(g(\rho)),
\end{equation}
where $f(\rho)$ is some unknown coefficient function to be determined,  $g(\rho)$ is a variable transformation and $\zeta(g)$ is to be so selected 
from the consideration of its fulfilling a second-order differential equation
\begin{equation}\label{de}
\zeta''(g(\rho))+Q_1(g)\zeta'(g(\rho))+R_1(g)\zeta(g(\rho))=0.
\end{equation}
such that the functions $Q_1(g)$ and $R_1(g)$ are well defined if a special function choice is made for it.
Inserting Eq. (\ref{extsol}) into Eq. (\ref{esrho}), we get 
\begin{equation}\label{ede}
\zeta''(g)+\bigg(\frac{2f'(\rho)}{f(\rho )g'(\rho)}+\frac{g''(\rho)}{g'(\rho)^2}+\frac{\tau}{\rho g'(\rho)}\bigg)\zeta'(g)
+\frac{1}{g'(\rho)^2}\bigg( \frac{f''(\rho)}{f(\rho)}+\frac{\tau f'(\rho)}{\rho f(\rho)}+2(E_{ext}-V_{ext})\bigg)\zeta(g)=0,
\end{equation}
where $V_{ext}=\frac{1}{2}\omega \rho^2+V_{new}$ and $\tau=\big(N+2s-1+ \lambda r(2N-r-1)\big)$.
On comparing Eq. (\ref{ede}) with Eq. (\ref{de}), we get
\begin{equation}\label{q}
Q_1(g)=\frac{2f'(\rho)}{f(\rho )g'(\rho)}+\frac{g''(\rho)}{g'(\rho)^2}+\frac{\tau}{\rho g'(\rho)}\\
~\mbox{and}\quad R_1(g)= \frac{1}{g'(\rho)^2}\bigg( \frac{f''(\rho)}{f(\rho)}+\frac{\tau f'(\rho)}{\rho f(\rho)}+2(E_{ext}-V_{ext})\bigg).
\end{equation}
After simplifying $Q_1(g)$, one finds that
\begin{equation}\label{f}
f(\rho)\simeq (g'(\rho))^{-\frac{1}{2}}\rho^{-\frac{\alpha}{2}}\exp\bigg(\frac{1}{2}\int^{g}Q_1(g)dg\bigg).
\end{equation}  
Using it in the expression of $R_1(g)$ we arrive at the form 
\begin{equation}
E_{ext}-V_{ext}=\frac{1}{2}\bigg[ \frac{g'''(\rho)}{2g'(\rho )}-\frac{3}{4}\frac{g''(\rho)^2}{g(\rho)^2}+\frac{\tau/2(\tau/2-1)}{\rho^2} + g'(\rho)^2\bigg(R_1(g)-\frac{Q_1'(g)}{2} - \frac{Q_1^2(g)}{4}\bigg)\bigg].
\end{equation}
which can be further simplified to 
\begin{equation}\label{e_vv}
E_{ext}-V_{ext}=\frac{1}{2} \bigg[\frac{\Delta V(\rho)}{2} +\frac{\tau/2(\tau/2-1)}{\rho^2} + g'(\rho)^2\bigg(R_1(g)-\frac{Q_1'(g)}{2} - \frac{Q_1^2(g)}{4}\bigg)\bigg].
\end{equation}
where the quantity $\Delta V(\rho)$ is the so-called Schwartzian
derivative,
\begin{equation}
  \Delta V(\rho) = \frac{g'''(\rho)}{g'(\rho )}-\frac{3}{2}\frac{g''(\rho)^2}{g(\rho)^2}.
  \label{eq:Delta}
\end{equation}
(\ref{e_vv}) is the main relation which we will exploit below in search for normalizable zero-energy solutions that reside in an extended class of truncated Calogero Sutherland model.
\par
\section{Normalizable zero-energy solutions}
As pointed out in Section $2$, the solutions of Eq. \eqref{eq:so(2,1)-cond} fall under three distinct categories as exemplified by the classes shown explicitly in Eqs. \eqref{eq:3cases}, \eqref{eq:3cases2} and \eqref{eq:3cases3}. Indeed these provide us with three unique realizations of $so(2,1)$ algebra. In the literature \cite{quesne1991} the matrix elements of the corresponding operators in the three types of system have been tied up in a single framework having different $so(2,1)$ coefficients. In an early attempt \cite{bagchi1997zero}, a class of QES potentials were generated for a class of QES rational potentials. This study was preceded by a systematic analysis undertaken by Daboul and Nieto \cite{daboul1994quantum, daboul1995exact, daboul1996exact} who addressed radial power law potentials to arrive at two interesting cases where the zero-energy wave function is normalizable with the corresponding state being either bound or unbound. In this connection, it is worth noting Barut's observation \cite{barut1980magnetic} that exact $E=0$ solutions could exist in relation to confinement (normalizable case) or leakage (non-normalizable case). In the context of an extended TCS model, our endeavour will be to seek normalizable zero-energy solutions related to it. As such, following the procedure of \cite{bagchi1997zero}, we first of all set $Q_1(g)=0$ which gives the  condition


\begin{equation}
    {f(\rho)}^2g'(\rho)\rho^\tau=\text{constant}
\end{equation}
This reduces equation (\ref{e_vv}) to the form
\begin{equation}\label{e_v}
E_{ext}-V_{ext}=\frac{\Delta V(\rho)}{4} +\frac{\tau/4(\tau/2-1)}{\rho^2} + \frac{1}{2}g'(\rho)^2R_1(g).
\end{equation}
Next, to get a meaningful representation of (\ref{e_v}) we put $R_1=E_T-V_T(g)$ and use the transformation $\rho=K
(g(\rho))$ to get
\begin{equation}
  E_T - V_T(g) = [E_{ext} - V_{ext}] 
  2 K'(\rho))^2 + \frac{1}{2} \Delta V(K(g))-\frac{\frac{\tau}{2}(\frac{\tau}{2}-1)}{K(g)^2}K'(g)^2
  \label{eq:E_T-V_T}
\end{equation}
 Now using $so(2,1)$ as potential algebra of the Schrodinger equation and putting $V_T= V_m$
\begin{equation}\label{eq:basic-eq}
  \left(\frac{1}{4} - m^2\right) \left(1 - F^2\right) - 2mFG + G^2 - E_T =
  2[K'(g)]^2 [V_{ext}(K(g)) - E_{ext}] - \frac{1}{2} \Delta V(K(g))+\frac{\frac{\tau}{2}(\frac{\tau}{2}-1)}{K(g)^2}K'(g)^2, 
\end{equation}
For clarity and simplicity, we have omitted the ``ext'' subscripts and will henceforth refer to $E$ and  $V$ without additional notation. Now taking first class of solution (\ref{eq:3cases}), using mapping function $\rho=K(g)={(e^g-1)}^{-1}$ and setting $E=0$ gives 
\begin{eqnarray}
  V(K(g)) & = & \left(1 - 4 m^2 + 4 b^2\right) \text{sech}~g ( \text{sech}~g - 2) - 8 m b (\tanh g~
           \text{sech}~g - 2 \tanh g + \sinh g) \nonumber \\
  & & \mbox{} - \left(4 E_T + \frac{1}{2}\right) \cosh g (\cosh g - 2)
           - 4 \left(E_T + m^2 - b^2-\frac{1}{8}\right) +\frac{\tau}{4}(\frac{\tau}{2}-1)(2\sinh^2 g+\sinh g)\nonumber \label{classI-prel}
\end{eqnarray}
 Thus by using the mapping function and above equation we can find $V(\rho)$
 \begin{equation}
  \mbox{(I)} \qquad V(\rho) = - \frac{A}{\left(2\rho^2 + 2\rho + 1\right)^2} - B \frac{2\rho + 1}{\rho
  (\rho+1) \left(2\rho^2 + 2\rho + 1\right)^2} + \frac{C}{\rho^2 (\rho+1)^2}-\frac{\frac{\tau}{4}(\frac{\tau}{2}-1)}{\rho^2},   \label{eq:classI} 
\end{equation}
with
\begin{equation}
  A = \frac{1}{2}(4 \left(m^2 - b^2\right) - 1), \qquad B = 2mb, \qquad C = -\frac{1}{2}( E_T+ \frac{1}{4}) =
  \left(k - \frac{1}{2}\right)^2 - \frac{1}{4} = k (k-1),.
\end{equation}
The first example of a QES potential with known $E=0$ is given in equation (\ref{eq:classI}). For nonnegative values of $C$ and $\tau=0$, its behaviour at the origin is similar to that of the (shifted) Coulomb effective potential $V(\rho) = \frac{\hbar^{2}}{2m}\frac{l(l+ 1)}{\rho^{2}} - \frac{e^{2}}{\rho}$.
The corresponding wave function for the first class potential with $m=k$ is $\psi(\rho)=\sqrt{K'(g(\rho))}\chi_0$, where $\chi_0 \sim G^{k-\frac{1}{2}}h$ is the ground state wave function of class I potential in Eq. (\ref{eq:3cases}) with $h=\exp{[b \tan^{-1}(\sinh{g})]}$ \cite{quesne1991} which gives 
\begin{equation}\label{29}
    \psi(\rho)=2^{k-\frac{1}{2}} \sqrt{\rho (\rho+1)} \rho^{-\frac{\tau }{2}} \left(\frac{b \rho (\rho+1)}{2 \rho (\rho+1)+1}\right)^{k-\frac{1}{2}} e^{b \tan^{-1} \left(\frac{2 \rho+1}{2 \rho^2+2 \rho}\right)}
\end{equation}
From the above expression, we can clearly see that for $\rho \to \infty$, the expression depends only on $\rho^{1-\frac{\tau}{2}}$. For the wavefunction to be well-behaved, the power must be negative. This imposes a constraint that $\tau > 2$. On the other hand, for $\rho \to 0$, the expression depends on $\rho^{k-\frac{\tau}{2}}$. For the wavefunction to be well-behaved in this case, the power $k-\frac{\tau}{2}$ must be positive, which imposes a constraint on $k$ such that $2k > \tau$. In summary, the wavefucntion is clearly well-behaved provided we put on restrictions $2k>\tau$ and $\tau>2$. Figure $1$ shows the plots of the probability density and wave function for different values of $k$, $\tau$, and $b$.

\begin{figure}[H]
     \centering
     \begin{subfigure}[b]{0.48\textwidth}
         \centering
         \includegraphics[width=7.5cm]{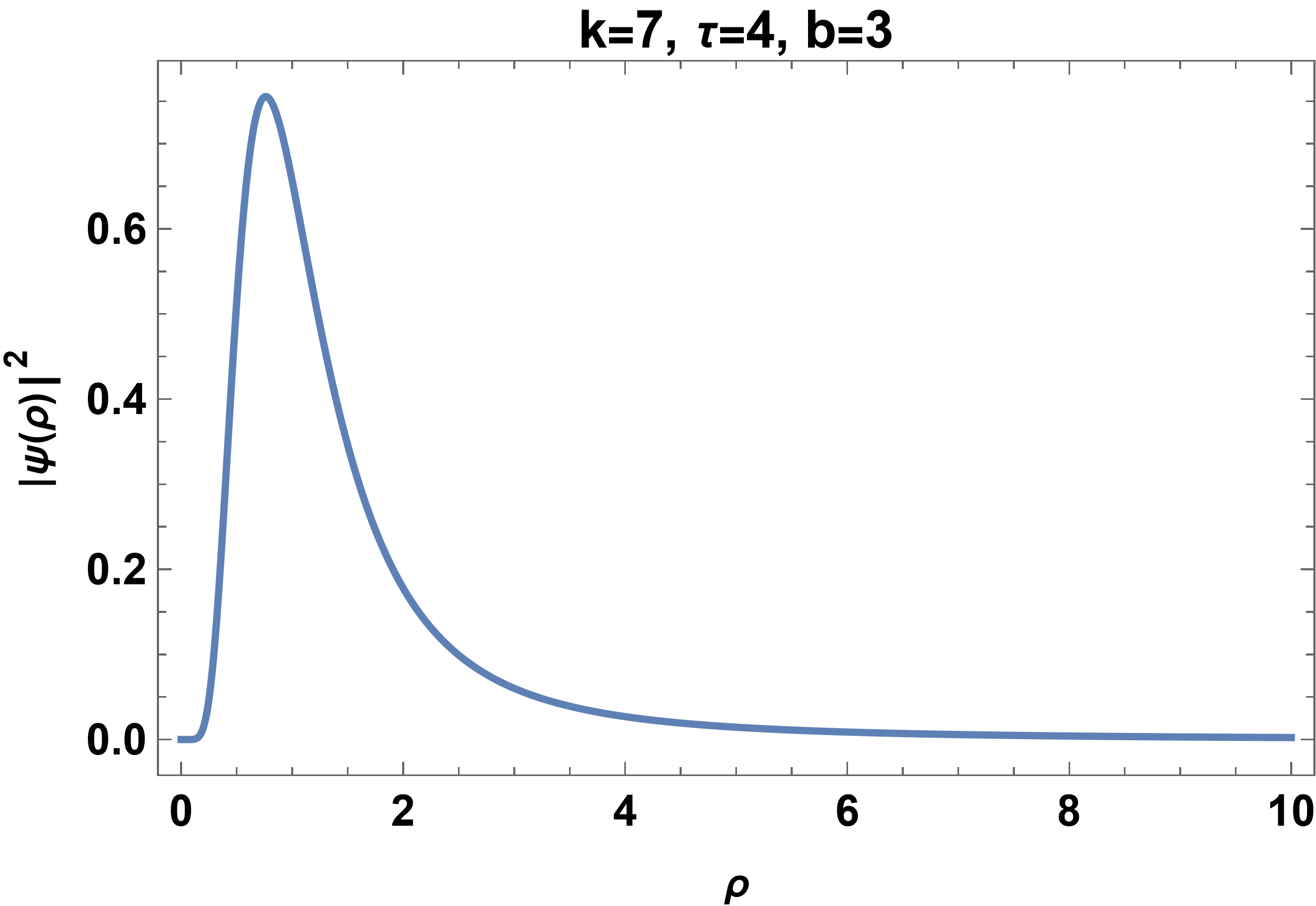}
         \caption{}
         \label{fig:1.1}
     \end{subfigure}
     \begin{subfigure}[b]{0.48\textwidth}
         \centering
         \includegraphics[width=7.5cm,]{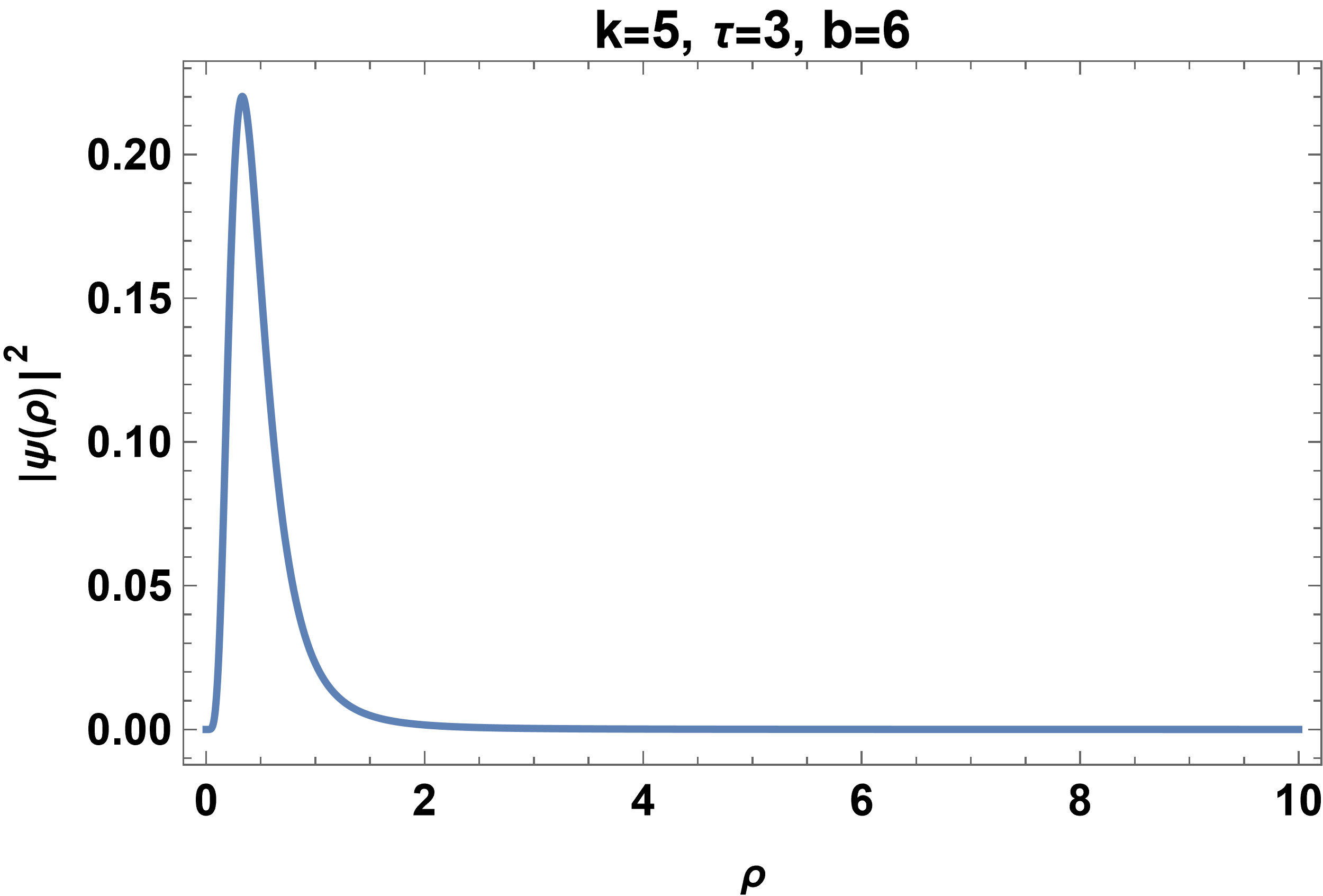}
         \caption{}
         \label{fig:1.2}
     \end{subfigure}
     \begin{subfigure}[b]{0.48\textwidth}
         \centering
         \includegraphics[width=7.5cm,height=5.5cm]{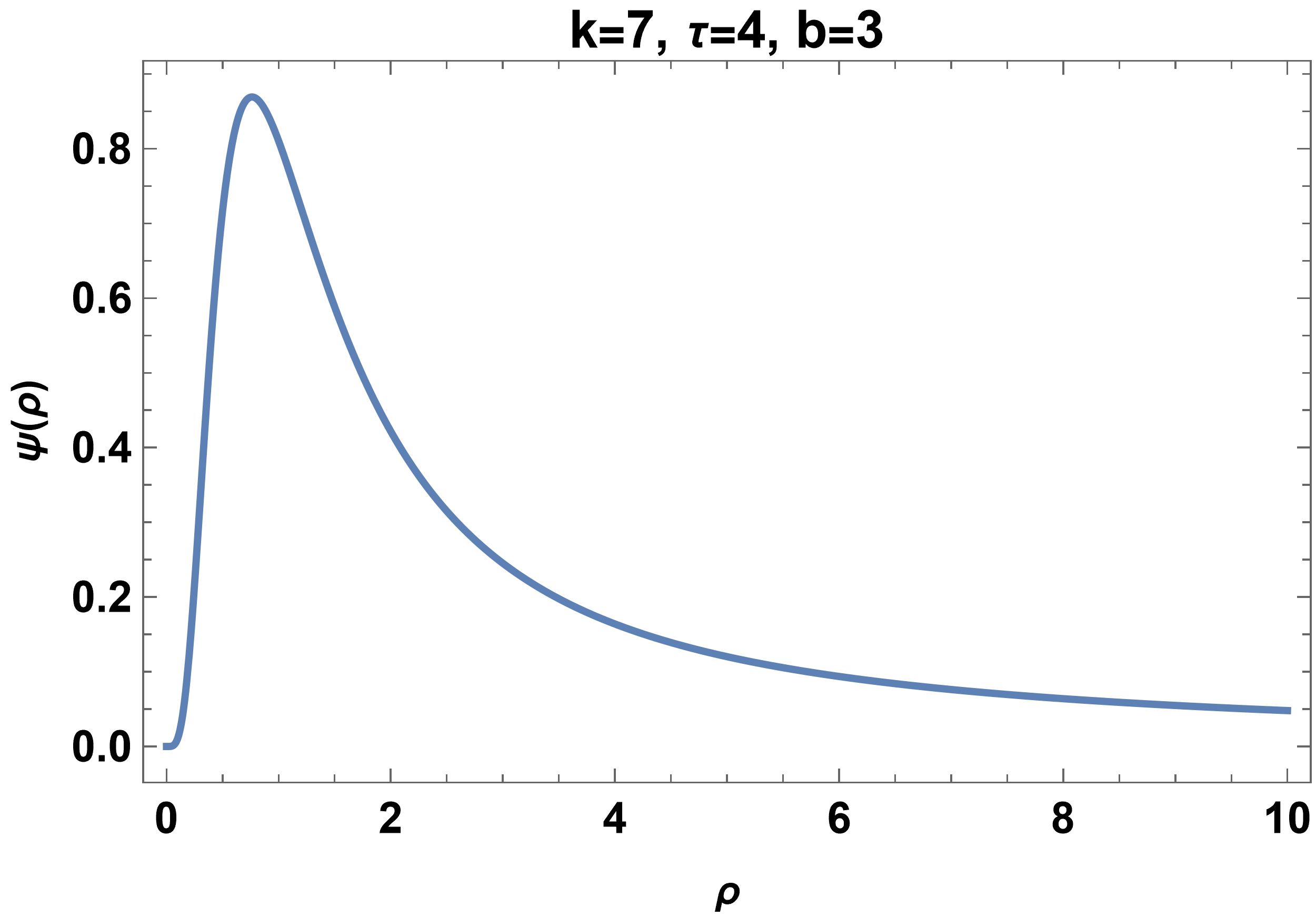}
         \caption{}
         \label{fig:32}
     \end{subfigure}
     \begin{subfigure}[b]{0.48\textwidth}
         \centering
         \includegraphics[width=7.5cm,height=5.5cm]{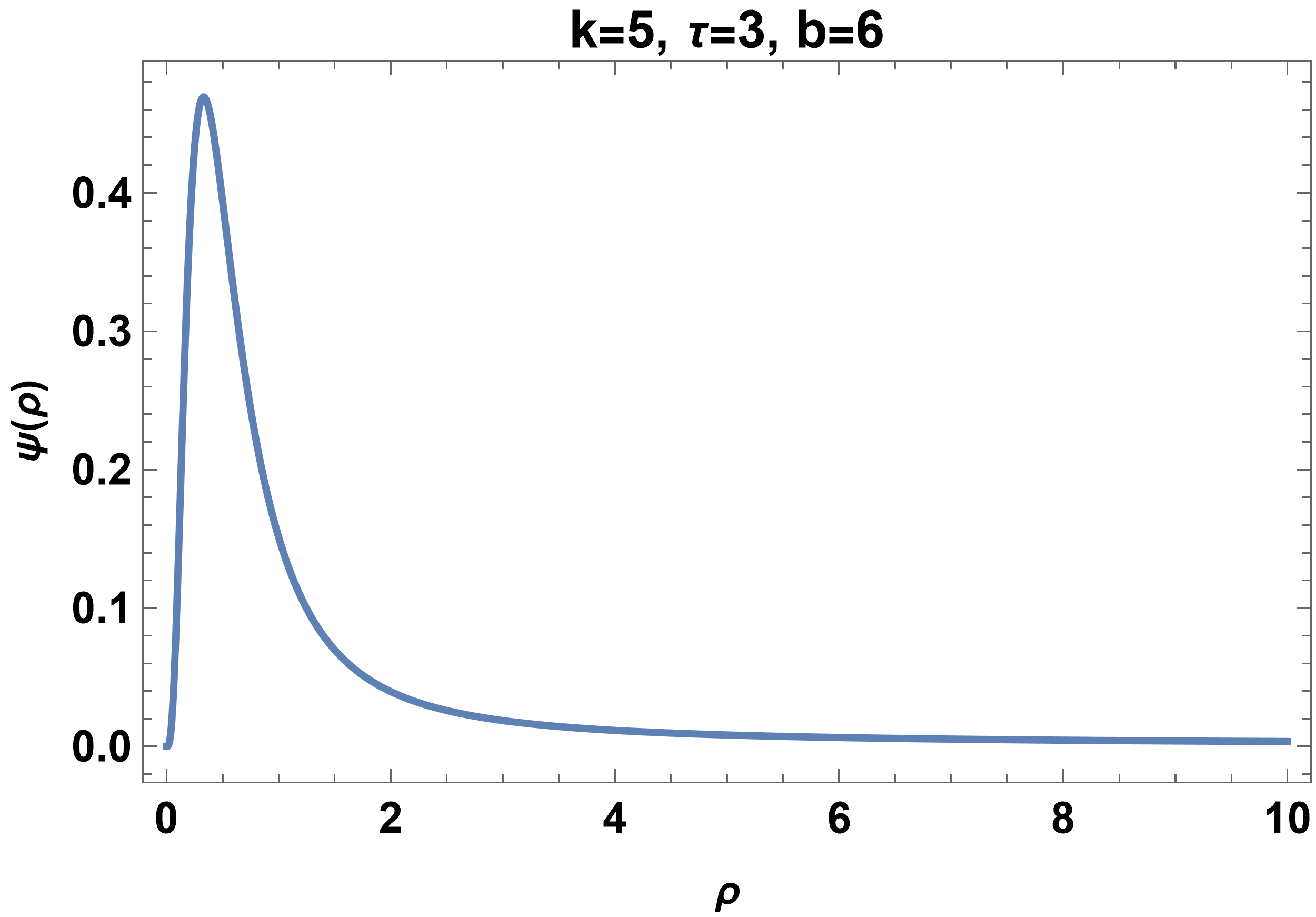}
         \caption{}
         \label{fig:38}
     \end{subfigure}
     \caption{(a) $\&$ (b) The plot illustrates the probability density for the first class potential at different values of $k$, $\tau$, and $b$ (c) $\&$ (d) The plot illustrates the wavefunction for the first class potential at different values of $k$, $\tau$, and $b$}.
     \label{fig: first class}
\end{figure}

\par
Similarly we can find the QES potential for other two class.
\begin{equation}\label{eq:classII}
     \mbox{(II)} \qquad V(\rho)= \frac{b^2}{2(\rho+1)^4}-\frac{mb}{\rho(\rho+1)^3}-\frac{E_T+\frac{1}{4}}{2\rho^2(\rho+1)^2}-\frac{\frac{\tau}{4}(\frac{\tau}{2}-1)}{\rho^2}
\end{equation}
The zero energy wave function for the second class potential with $m=k$ is $\psi(\rho)=\sqrt{K'(g(\rho))}\chi_0$, where $\chi_0 \sim G^{k-\frac{1}{2}}h$ with $h=\exp{[-be^{-g}]}$ which gives 
\begin{equation}\label{31}
    \psi(\rho)=\sqrt{\rho (\rho+1)} e^{-\frac{b \rho}{\rho+1}} \rho^{-\frac{\tau }{2}} \left(\frac{b \rho}{\rho+1}\right)^{k-\frac{1}{2}}
\end{equation}
It is clear that $ \psi(\rho)$ has an acceptable asymptotic behaviour provided the conditions $2k>\tau$, $\tau\geq 4$,  and $b>0$ are met. A graphical illustration of the probability density and the wavefunction is shown in Figure $2$ for different values of $k$, $\tau$, and $b$. 

\begin{figure}[H]
     \centering
     \begin{subfigure}[b]{0.48\textwidth}
         \centering
         \includegraphics[width=7.5cm]{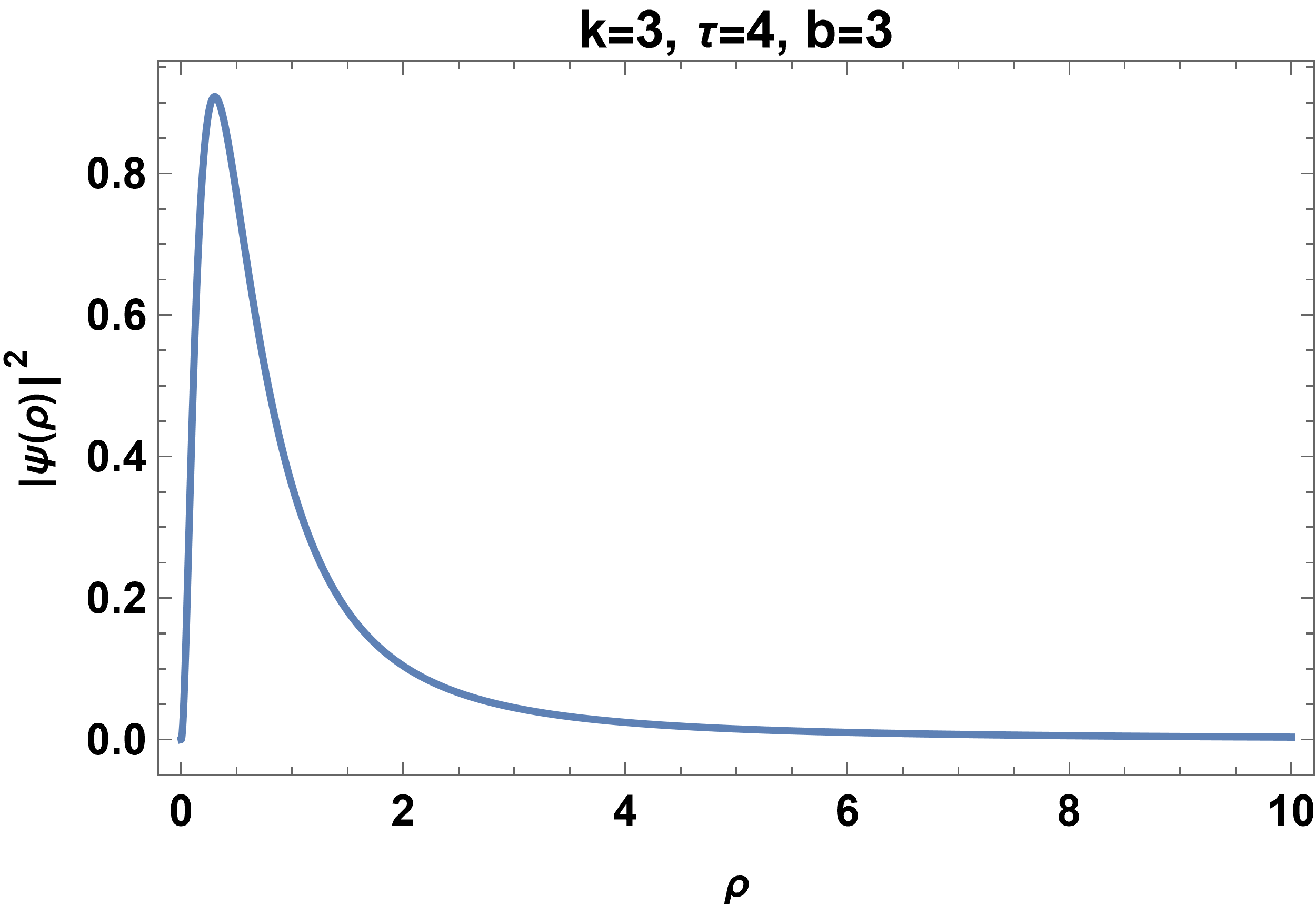}
         \caption{}
         \label{fig:2.1}
     \end{subfigure}
     \begin{subfigure}[b]{0.48\textwidth}
         \centering
         \includegraphics[width=7.5cm,]{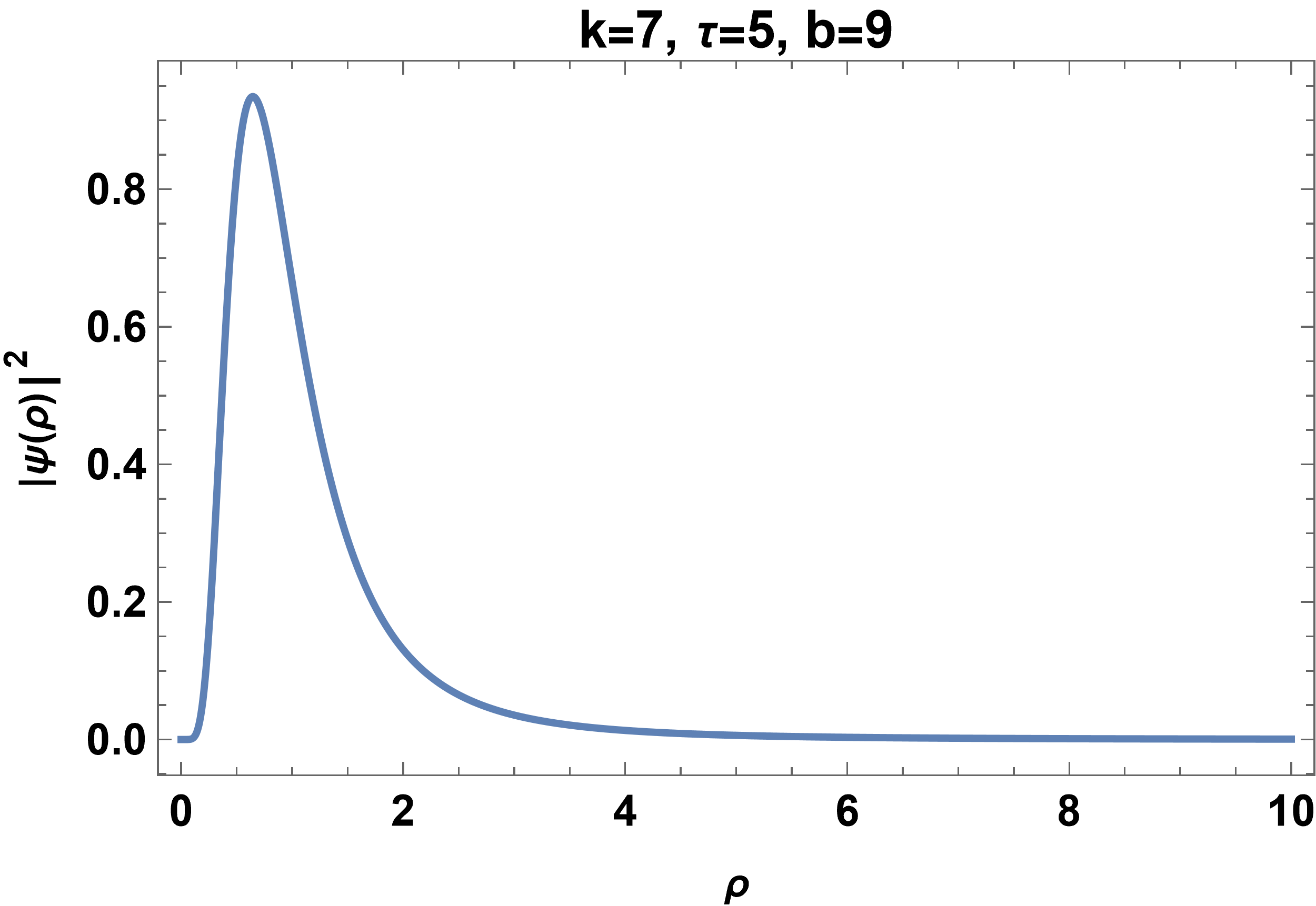}
         \caption{}
         \label{fig:2.2}
     \end{subfigure}
     \begin{subfigure}[b]{0.48\textwidth}
         \centering
         \includegraphics[width=7.5cm,height=5.5cm]{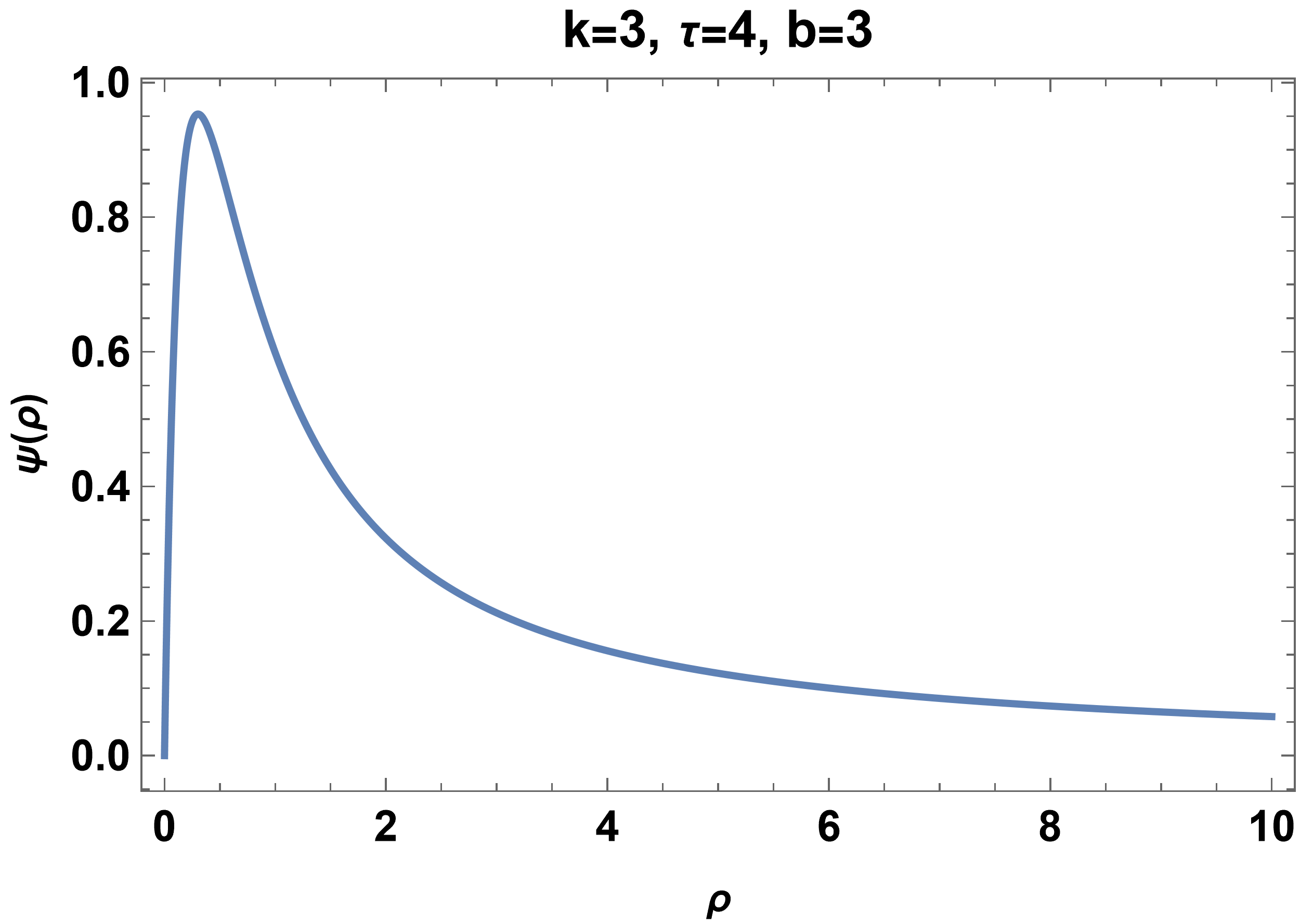}
         \caption{}
         \label{fig:2.3}
     \end{subfigure}
     \begin{subfigure}[b]{0.48\textwidth}
         \centering
         \includegraphics[width=7.5cm,height=5.5cm]{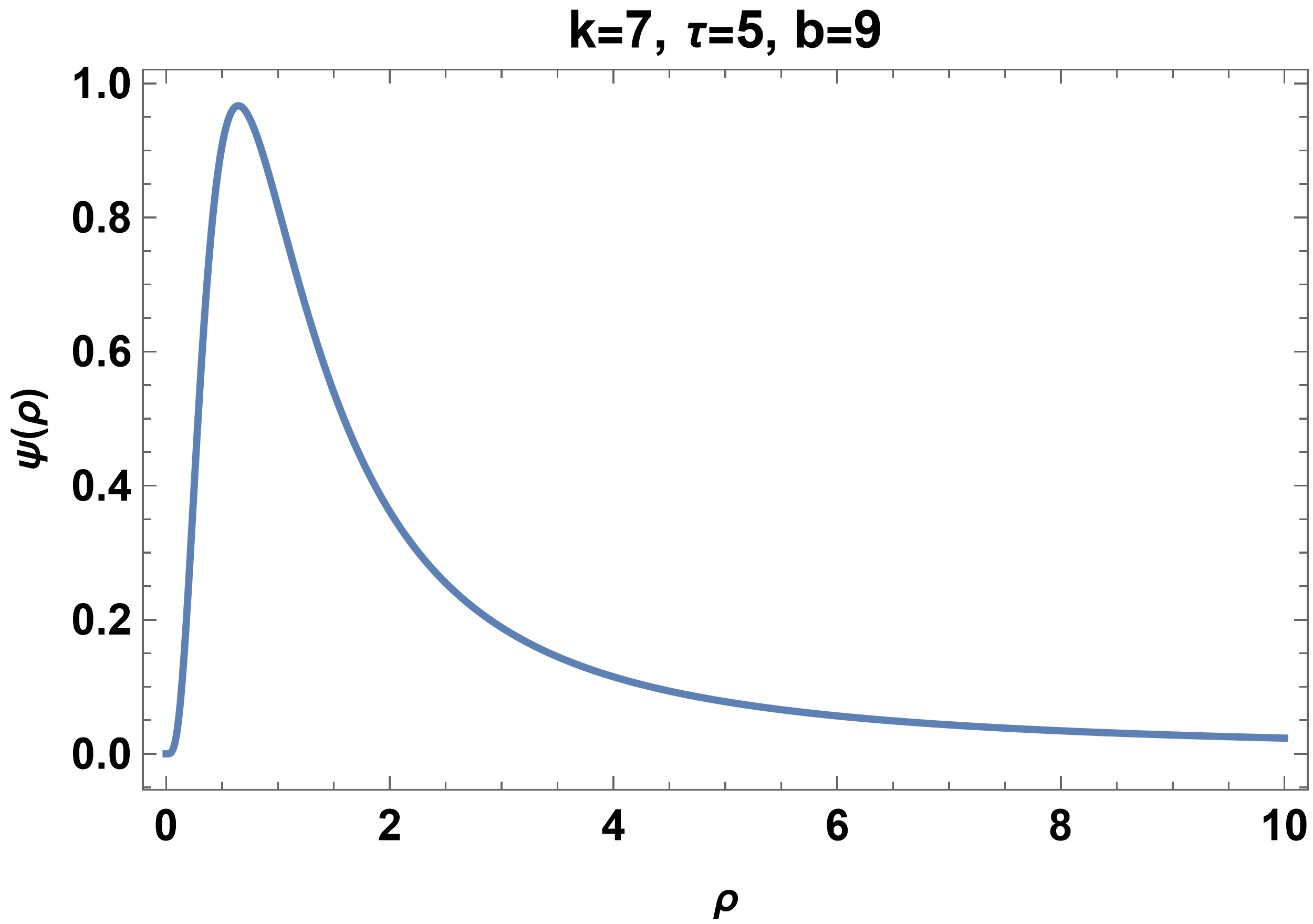}
         \caption{}
         \label{fig:2.4}
     \end{subfigure}
     \caption{(a) $\&$ (b) The plot illustrates the probability density for the second class potential at different values of $k$, $\tau$, and $b$ (c) $\&$ (d) The plot illustrates the wavefunction for the second class potential at different values of $k$, $\tau$, and $b$}.
     \label{fig: second class}
\end{figure}

The third kind of QES potential has the form 
 \begin{equation}\label{eq:classIII}
     \mbox{(III)} \qquad V(\rho)= \frac{4(m+b)^2-1}{2(2\rho+1)^2}-\frac{2mb}{\rho(\rho+1)}-\frac{E_T+\frac{1}{4}}{2\rho^2(\rho+1)^2}-\frac{\frac{\tau}{4}(\frac{\tau}{2}-1)}{\rho^2}
 \end{equation}
Its supporting wave function with $m=k$ is given by 
\begin{equation}\label{33}
    \psi(\rho)=2^{k-\frac{1}{2}} \sqrt{\rho (\rho+1)} \left(\frac{1}{2 \rho+1}\right)^b \rho^{-\frac{\tau }{2}} \left(\frac{b \rho (\rho+1)}{2 \rho+1}\right)^{k-\frac{1}{2}}
\end{equation}
{The normalization integral is given by}

{\begin{equation}
  \int_{0}^{\infty}{|\psi|}^2 d\rho =\int_{0}^{\infty} \frac{\rho^{-\tau}[\rho(\rho+1)]^{2k}}{(2\rho+1)^{2k+2b-1}} d\rho
\end{equation}
{To examine its convergence, we apply the criterion which states that if $\lim_{\rho \to \infty} \rho^\alpha f(\rho) = L \neq 0$ and $\alpha \leq 1$, then $\int_{a}^{\infty} f(\rho) d\rho$ diverges. In this case, as $\rho \to \infty$, $f(\rho) \sim \rho^{2k-2b-\tau+1}$. Setting $\alpha = 2b-2k+\tau-1$, we find that $\lim_{\rho \to \infty} \rho^{\alpha} f(\rho) = L \neq 0$. The integral converges if $2b-2k+\tau-1 > 1$ or $b > k - \frac{\tau}{2} + 1$, and diverges if $b \leq k - \frac{\tau}{2} + 1$.

It can seen to be well-behaved when the conditions $2k>\tau$ and $b>k-\frac{\tau}{2}+1$ hold. The plots of the probability density and the wavefunction are shown in Figure $3$ for different values of $k$, $\tau$, and $b$. Representative graphs of the potential for different choices of the parameters are sketched in Figure $4$. It is interesting to note here that unlike in \cite{bagchi1997zero} where, of the three classes of solutions admitted by $so(2,1)$, only one yielded to normalizable zero-energy wave function provided certain convergence condition was satisfied, here all the threes cases support normalizable zero-energy solutions subject to certain constraints on the coupling parameters. In fact this is a distinctive feature of the Casimir of a suitably mapped rationally extended TCS model under a point canonical transformation.

\section{Summary}
To summarize, we have dealt with three cases corresponding to the restrictions on the guiding functions of so$(2,1)$. An important outcome of our analysis is that in all the three cases we found normalizable wavefunctions by placing plausible restrictions on the coupling parameters. This is in marked contrast to a similar situation one faced in a previous work \cite{bagchi1997zero} while addressing  a class of QES rational potentials with known $E = 0$ eigenvalue. It was found that subject to certain
convergence condition being satisfied, the wave function of the zero-energy state for only one class of potentials turned out to be normalizable. Addressing a rationally extended truncated Calogero-Sutherland model, we subjected the associated Schr\"{o}dinger equation to a point canonical transformation and matched the resulting energy expression with the Casimir of the so$(2,1)$ algebra. This gave three new classes of QES rational potentials corresponding to the $E=0$ state. For all of
them, we showed that the solutions for the eigenvalue equation at zero energy were normalizable and asymptotically smooth provided the coupling parameters were properly restricted. 
Concerning the normalizability of the wave functions appearing in (\ref{29}), (\ref{31}) and (\ref{33}), it may be pointed out that these are the projected versions of the original form (\ref{sch}) conforming to the three different cases of the so(2,1) algebra with regard to the solutions (\ref{eq:3cases}), (\ref{eq:3cases2}), and (\ref{eq:3cases3} ) respectively. Indeed the three types of solutions arise as a result of the factorization of the general structure (\ref{sch}) that facilitates tractability of the wavefunction in a straightforward way. We have treated the normalizability problem individually depending on the restrictions of the F and G functions rather than handling (\ref{sch}) in a general framework which would result in unavoidable complications.

\begin{figure}[H]
     \centering
     \begin{subfigure}[b]{0.48\textwidth}
         \centering
         \includegraphics[width=7.5cm]{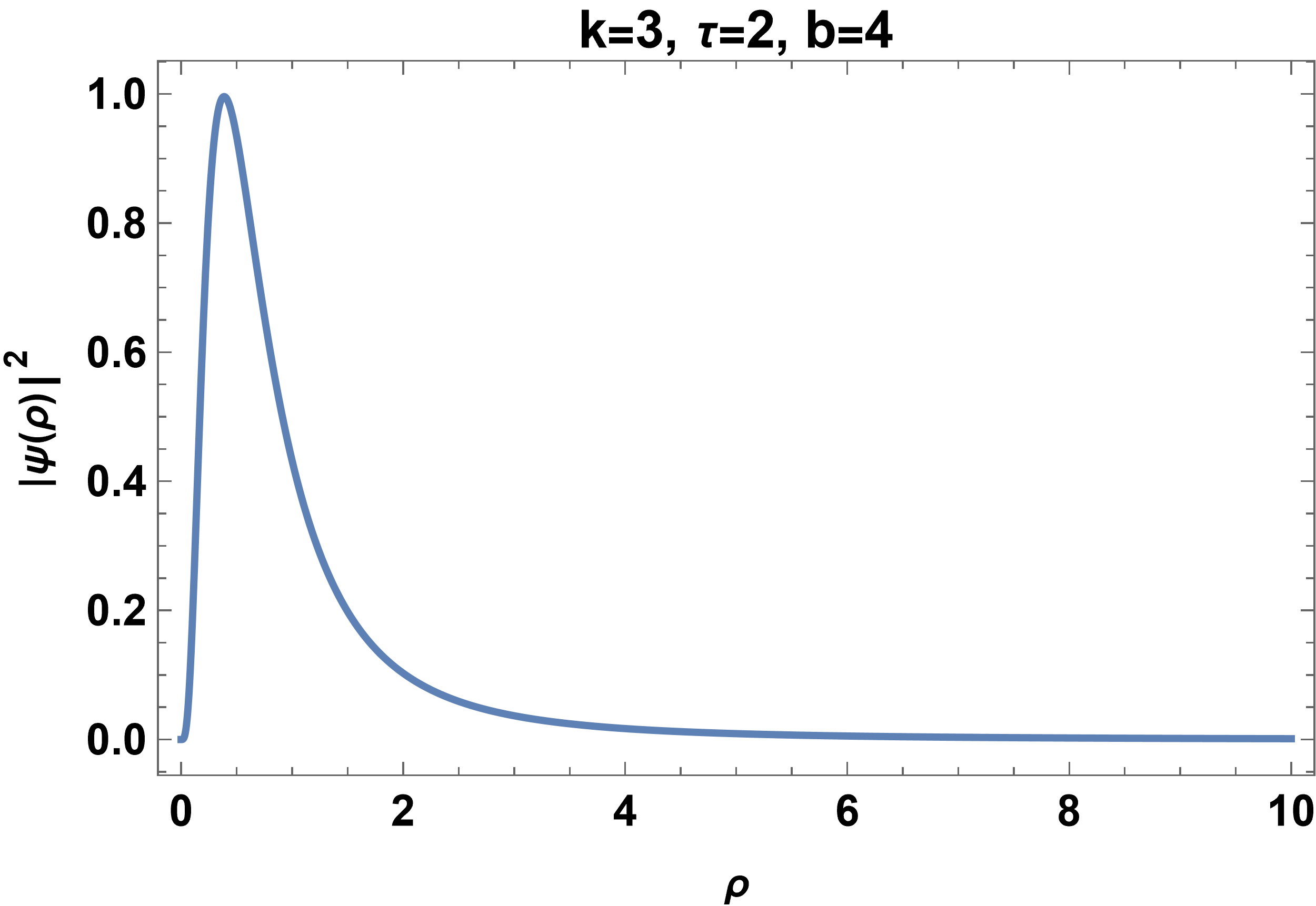}
         \caption{}
         \label{fig:3.1}
     \end{subfigure}
     \begin{subfigure}[b]{0.48\textwidth}
         \centering
         \includegraphics[width=7.5cm,]{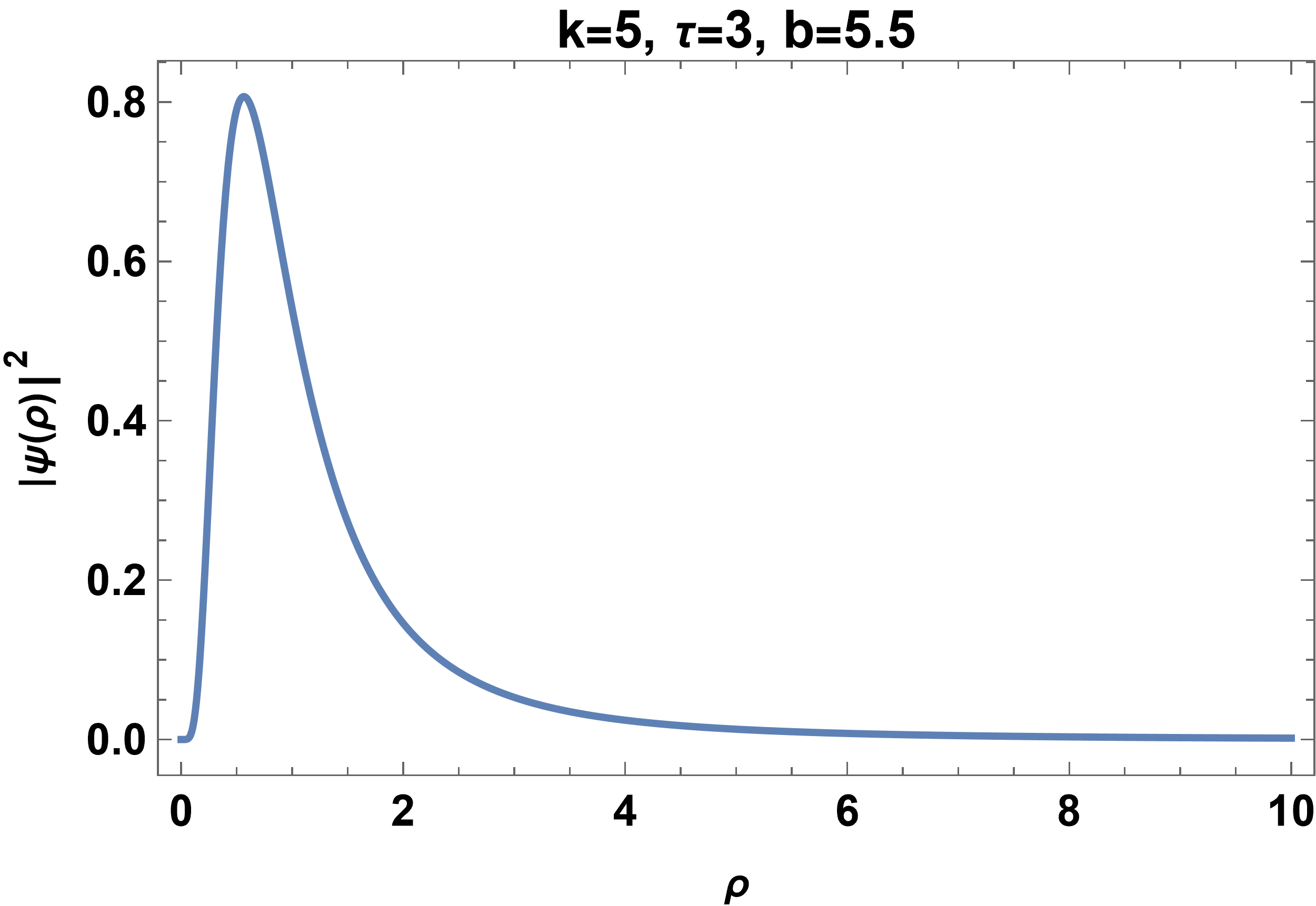}
         \caption{}
         \label{fig:3.2}
     \end{subfigure}
     \begin{subfigure}[b]{0.48\textwidth}
         \centering
         \includegraphics[width=7.5cm,height=5.5cm]{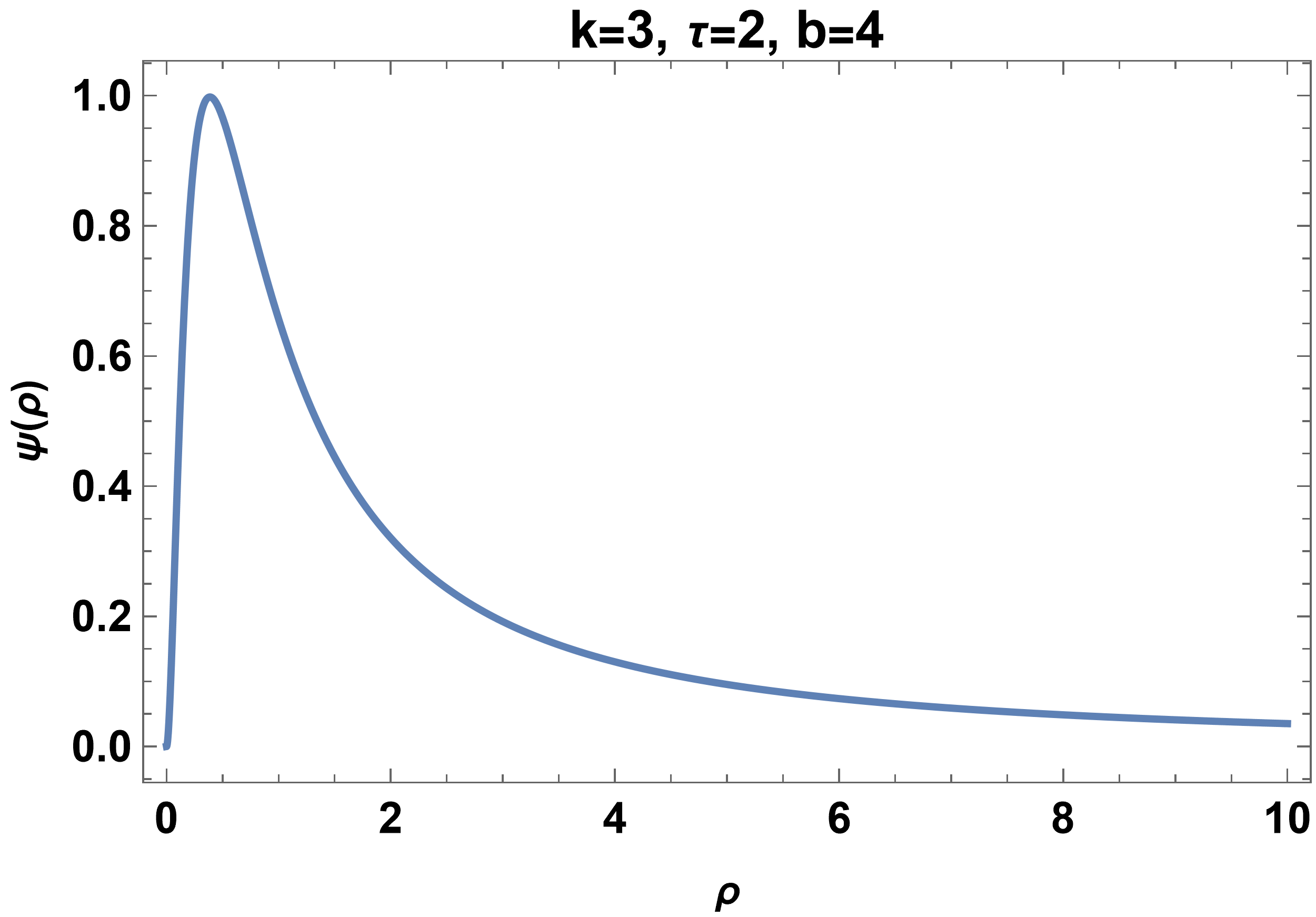}
         \caption{}
         \label{fig:3.3}
     \end{subfigure}
     \begin{subfigure}[b]{0.48\textwidth}
         \centering
         \includegraphics[width=7.5cm,height=5.5cm]{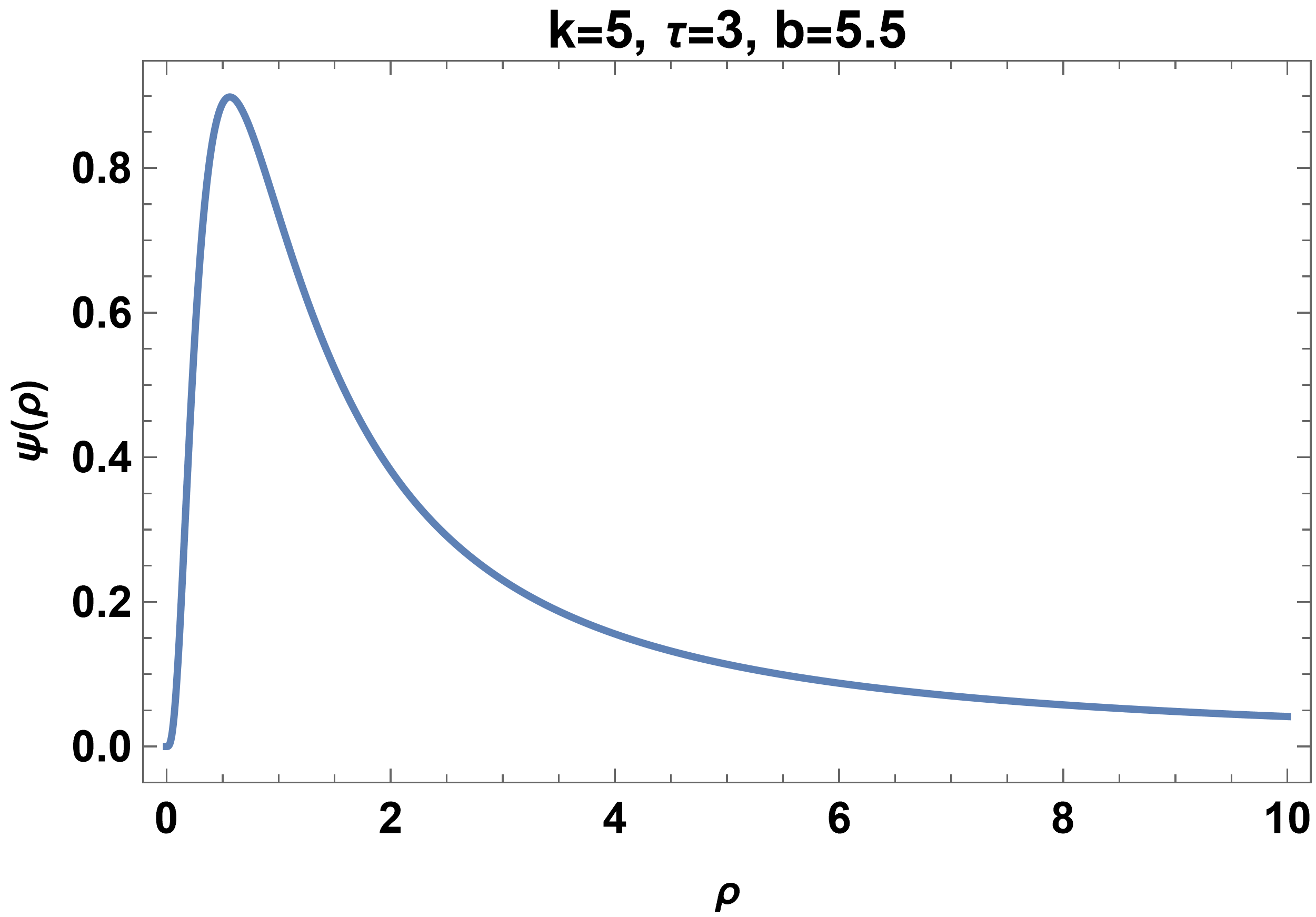}
         \caption{}
         \label{fig:3.4}
     \end{subfigure}
     \caption{(a) $\&$ (b) The plot illustrates the probability density for the third class potential at different values of $k$, $\tau$, and $b$ (c) $\&$ (d) The plot illustrates the wavefunction for the third class potential at different values of $k$, $\tau$, and $b$}.
     \label{fig: third class}
\end{figure}

\begin{figure}[H]
     \centering
     \begin{subfigure}[b]{0.48\textwidth}
         \centering
         \includegraphics[width=7.5cm]{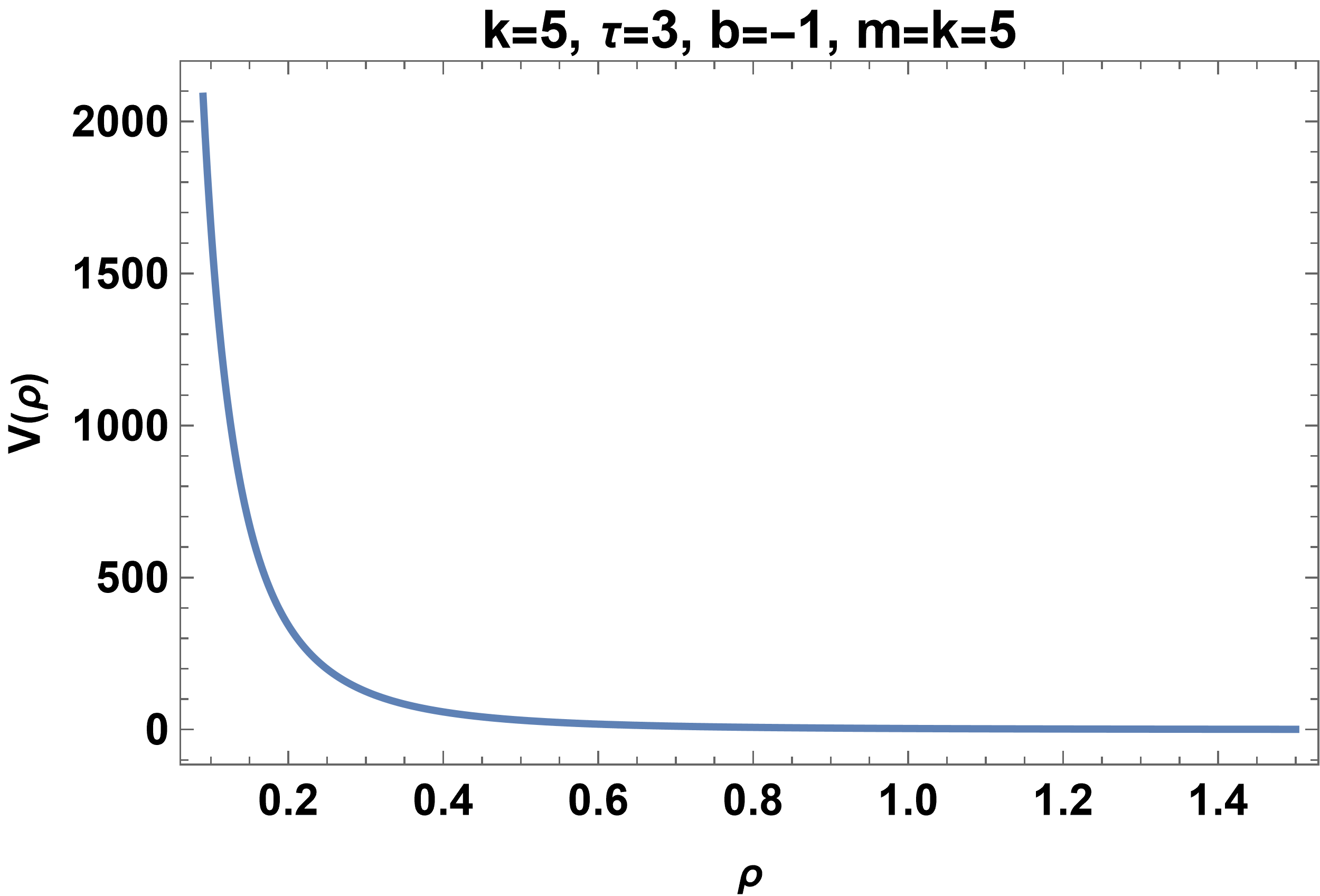}
         \caption{}
         \label{fig:v1.1}
     \end{subfigure}
     \begin{subfigure}[b]{0.48\textwidth}
         \centering
         \includegraphics[width=7.5cm,]{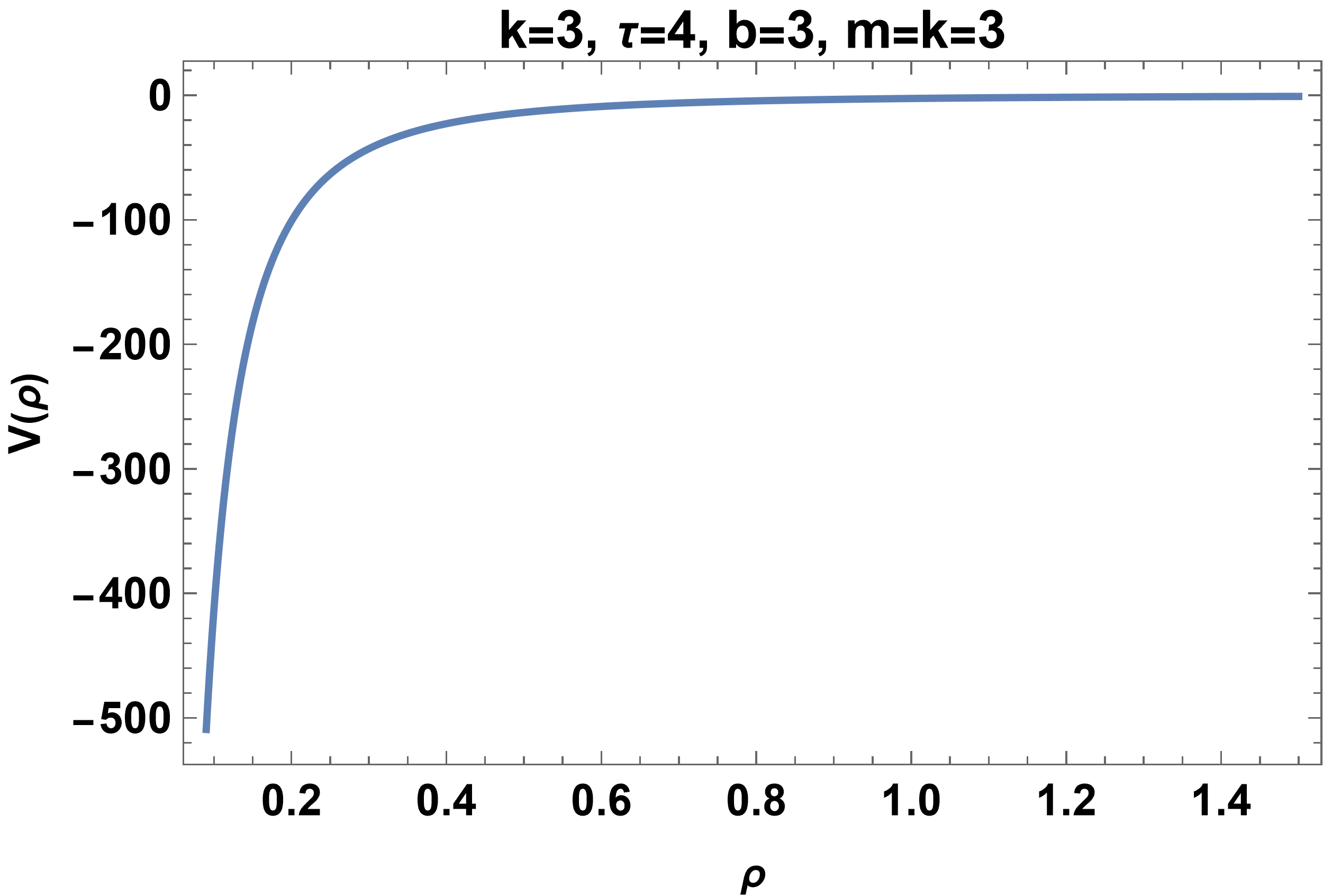}
         \caption{}
         \label{fig:v2.1}
     \end{subfigure}
     \begin{subfigure}[b]{0.48\textwidth}
         \centering
         \includegraphics[width=7.5cm,height=5.5cm]{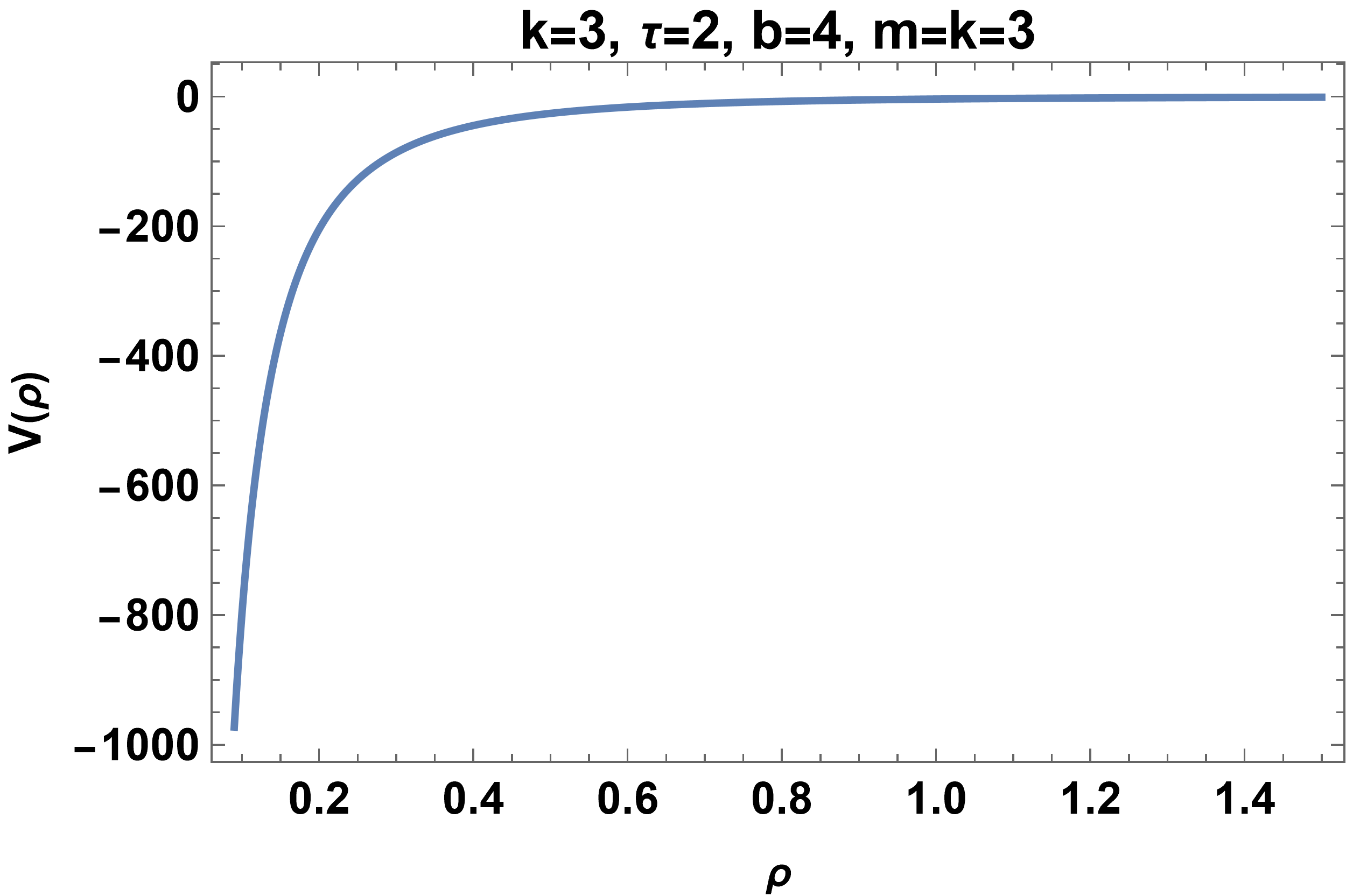}
         \caption{}
         \label{fig:v3.1}
     \end{subfigure}
     \caption{(a) The plots of potential $V(\rho)$ of equation (\ref{eq:classI}). (b) The plot of potential $V(\rho)$ of equation (\ref{eq:classII}). (c) The plot of potential $V(\rho)$ of equation (\ref{eq:classIII}).}.
     \label{fig: potential plots}
\end{figure}




\section{Acknowledgements}
 BPM acknowledges the research grant for faculty under the IoE Scheme (Number 6031) of Banaras Hindu University. BB is grateful to Brainware University for infrastructural support.

\section{Data availability statement}

All data supporting the findings of this study are included in the article.
 
\bibliographystyle{elsarticle-num}
\bibliography{Ref}

\begin{thebibliography}{10}
\expandafter\ifx\csname url\endcsname\relax
  \def\url#1{\texttt{#1}}\fi
\expandafter\ifx\csname urlprefix\endcsname\relax\def\urlprefix{URL }\fi
\expandafter\ifx\csname href\endcsname\relax
  \def\href#1#2{#2} \def\path#1{#1}\fi

\bibitem{ito1970partantipart}
H.~Ito, Zero-energy bound states of two dirac particles: On the properties of
  eigenvalue spectra of the o(4) families, Progress of Theoretical Physics 43
  (1970) 1035.

\bibitem{daboul1994quantum}
J.~Daboul, M.~M. Nieto, Quantum bound states with zero binding energy, Physics
  Letters A 190~(5-6) (1994) 357--362.

\bibitem{daboul1995exact}
J.~Daboul, M.~M. Nieto, Exact, e= 0, classical solutions for general power-law
  potentials, Physical Review E 52~(4) (1995) 4430.

\bibitem{daboul1996exact}
J.~Daboul, M.~M. Nieto, Exact, e= 0, quantum solutions for general power-law
  potentials, International Journal of Modern Physics A 11~(20) (1996)
  3801--3817.

\bibitem{Alhaidari}
A.~D. Alhaidari, Exact solutions of dirac and schr\"odinger equations for a
  large class of power-law potentials at zero energy, International Journal of
  Modern Physics A 17 (2002) 4511--4566.

\bibitem{Halberg}
A.~Schulze-Halberg, Closed-form solutions of the schr\"odinger equation for a
  particle on the torus, Foundations of Physics Letters A 15 (2002) 585--589.

\bibitem{Makowski}
A.~Makowski, K.~Gorska, Unusual properties of some e= 0 localized states and
  the quantum-classical correspondence, Physics Letters A 362~(1) (2007)
  26--30.

\bibitem{Kaleta}
K.~Kaleta, J.~L{\H{o}}rinczi, Zero-energy bound state decay for non-local
  schr{\"o}dinger operators, Communications in Mathematical Physics 374~(3)
  (2020) 2151--2191.

\bibitem{barut1980magnetic}
A.~O. Barut, Magnetic resonances between massive and massless spin-1/2
  particles with magnetic moments, Journal of Mathematical Physics 21~(3)
  (1980) 568--570.

\bibitem{bagchi1997zero}
B.~Bagchi, C.~Quesne, Zero-energy states for a class of quasi-exactly solvable
  rational potentials, Physics Letters A 230~(1-2) (1997) 1--6.

\bibitem{shifmanqes}
M.~A. Shifman, New findings in quantum mechanics (partial algebraization of the
  spectral problem), International Journal of Modern Physics A 4 (1989)
  2897--2952.

\bibitem{khare1995}
F.~Cooper, A.~Khare, U.~Sukhatme, Supersymmetry and quantum mechanics, Physics
  Reports 251 (1995) 267.

\bibitem{fernandez}
D.~J. Fernandez~C, Trends in supersymmetric quantum mechanics, Integrability,
  Supersymmetry and Coherent States, Springer (2019) 37--68.

\bibitem{levai1989}
G.~L\'{e}vai, A search for shape-invariant solvable potentials, Journal of
  Physics A: Mathematical and General 22 (1989) 689.

\bibitem{alhassid1}
Y.~Alhassid, F.~G\"{u}rsey, F.~Iachello, Potential scattering, transfer matrix,
  and group theory, Physical Review Letters 50 (1983) 873.

\bibitem{alhassid2}
Y.~Alhassid, F.~G\"{u}rsey, F.~Iachello, Group theory approach to scattering,
  Annals of Physics 148 (1983) 346.

\bibitem{alhassid3}
J.~Wu, Y.~Alhassid, The potential group approach and hypergeometric
  differential equations, Journal of Mathematical Physics 31~(3) (1990)
  557--562.

\bibitem{quesne1991}
M.~Engelfield, C.~Quesne, Dynamical potential algebras for gendenshtein and
  morse potentials, Journal of Physics A: Mathematical and General 24~(15)
  (1991) 3557.

\bibitem{yadav2015group}
R.~K. Yadav, N.~Kumari, A.~Khare, B.~P. Mandal, Group theoretic approach to
  rationally extended shape invariant potentials, Annals of Physics 359 (2015)
  46--54.

\bibitem{ramos2017short}
A.~Ramos, B.~Bagchi, A.~Khare, N.~Kumari, B.~P. Mandal, R.~K. Yadav, A short
  note on ``group theoretic approach to rationally extended shape invariant
  potentials"[ann. phys. 359 (2015) 46--54], Annals of Physics 382 (2017)
  143--149.

\bibitem{turbinerqes1}
A.~V. Turbiner, A.~G. Ushveridze, Spectral singularities and quasi-exactly
  solvable quantal problem, Physics Letters A 126 (1987) 181--183.

\bibitem{turbinerqes2}
A.~V. Turbiner, Quasi-exactly-solvable problems and sl(2) algebra,
  Communications for Mathematical Physics 118 (1988) 467.

\bibitem{ushveridzebook}
A.~G. Ushveridze, Quasi-exactly solvable models in quantum mechanics, Institute
  of Publishing, Bristol, Institute of Physics Publishing, Bristol, 1994.

\bibitem{bender1996}
C.~M. Bender, G.~V. Dunne, Quasi-exactly solvable systems and orthogonal
  polynomials, Journal of Mathematical Physics 37 (1996) 6--11.

\bibitem{tkachuk}
V.~M. Tkachuk, Quasi-exactly solvable potentials with two known eigenstates,
  Physics Letters A 245 (1998) 177.

\bibitem{brihaye}
Y.~Brihaye, N.~Debergh, J.~Ndimubandi, On a lie algebraic approach of
  quasi-exactly solvable potentials with two known eigenstates, Modern Physics
  Letters A 16~(19) (2001) 1243--1251.

\bibitem{fring2019}
A.~Fring, T.~Frith, Quasi-exactly solvable quantum systems with explicitly
  time-dependent hamiltonians, Physics Letters A 383 (2019) 158--163.

\bibitem{khare1998quasi}
A.~Khare, B.~P. Mandal, Do quasi-exactly solvable systems always correspond to
  orthogonal polynomials?, Physics Letters A 239~(4-5) (1998) 197--200.

\bibitem{khare2009new}
A.~Khare, B.~P. Mandal, New quasi-exactly solvable hermitian as well as
  non-hermitian-invariant potentials, Pramana 73~(2) (2009) 387--395.

\bibitem{basu2017quasi}
B.~Basu-Mallick, B.~P. Mandal, P.~Roy, Quasi exactly solvable extension of
  calogero model associated with exceptional orthogonal polynomials, Annals of
  Physics 380 (2017) 206--212.

\bibitem{dutraces1}
A.~de~Souza~Dutra, H.~Boschi-Filho, so(2,1) lie algebra and the green's
  functions for the conditionally exactly solvable potentials, Phys. Rev. A 50
  (1994) 2915.

\bibitem{dutraces2}
A.~de~Souza~Dutra, Conditionally exactly soluble class of quantum potentials,
  Phys. Rev. A 47 (1993) R2435.

\bibitem{grosche1}
C.~Grosche, Conditionally solvable path integral problems, Journal of Physics
  A: Mathematical and General 28 (1995) 5889.

\bibitem{grosche2}
C.~Grosche, Conditionally solvable path integral problems: Ii. natanzon
  potentials, Journal of Physics A: Mathematical and General 29 (1996) 365.

\bibitem{duttces}
R.~Dutt, A.~Khare, Y.~Varshni, New class of conditionally exactly solvable
  potentials in quantum mechanics, Journal of Physics A: Mathematical and
  General 28~(3) (1995) L107.

\bibitem{znojil1}
M.~Znojil, Comment on conditionally exactly soluble class of quantum
  potentials, Physical Review A 61 (2000) 066101.

\bibitem{znojil2}
R.~Roychoudhury, P.~Roy, M.~Znojil, G.~L{\'e}vai, Comprehensive analysis of
  conditionally exactly solvable models, Journal of Mathematical Physics 42~(5)
  (2001) 1996--2007.

\bibitem{zaslavskii1990qes}
O.~B. Zaslavskii, Quasi-exactly solvable problems and su(1, 1) group, Modern
  Physics Letters A 9 (1994) 1501--1505.

\bibitem{bagchi2004ces}
B.~Bagchi, C.~Quesne, Conditionally exactly solvable potential and dual
  transformation in quantum mechanics, Journal of Physics A: Mathematical and
  General 37 (2004) L133--L135.

\bibitem{pittman2017truncated}
S.~Pittman, M.~Beau, M.~Olshanii, A.~del Campo, Truncated calogero-sutherland
  models, Physical Review B 95~(20) (2017) 205135.

\bibitem{yadav2019rationally}
R.~K. Yadav, A.~Khare, N.~Kumari, B.~P. Mandal, Rationally extended many-body
  truncated calogero--sutherland model, Annals of Physics 400 (2019) 189--197.

\bibitem{Jain}
S.~R. Jain, A.~Khare, An exactly solvable many-body problem in one dimension
  and the short-range dyson model, Physics Letters A 262~(1) (1999) 35--39.

\bibitem{Calogero}
F.~Calogero, Solution of the one-dimensional n-body problems with quadratic
  and/or inversely quadratic pair potentials, Journal of Mathematical Physics
  12 (1971) 419--436.

\bibitem{Sutherland}
B.~Sutherland, Quantum many-body problem in one dimension: Ground state,
  Journal of Mathematical Physics 12 (1971) 246--250.

\bibitem{Sudarshan}
A.~Bhattacharjie, E.~Sudarshan, A class of solvable potentials, Il Nuovo
  Cimento Series 10 25 (1962) 864--879.

\end{thebibliography}

\end{document}